\documentclass[amsmath,amssymb,aps,prx,twocolumn,superscriptaddress,showpacs]{revtex4-2} 
\usepackage{amssymb}
\usepackage{suffix}
\usepackage{mathtools}
\usepackage[utf8]{inputenc}
\usepackage{booktabs}
\usepackage{cases}
\usepackage[multiple]{footmisc}
\usepackage{dcolumn}
\usepackage{color,soul}
\usepackage{rotating}
\usepackage{perpage}
\usepackage{xcolor}
\usepackage{soul}
\usepackage{tikz}
\usepackage[T1]{fontenc}
\usepackage{etoolbox}
\usepackage{graphics}
\usepackage{siunitx}
\usepackage{hyperref}
\usepackage{float}
\usepackage{multirow}
\usepackage{mathtools}
\usepackage{bm}
\usepackage{url}

\usepackage{tikz}
\usepackage{tikz-3dplot}
\usepackage [outdir=./]{epstopdf}
\usepackage{setspace}
\begin{document}
\title{Light-induced switch based on edge modes in irradiated thin topological insulators}

\author{Zahra Askarpour}
\address{Physics Department, Iran University of Science and Technology, P. O. Box, 16844, Narmak, Tehran, Iran}
\author{Hosein Cheraghchi}
\email{cheraghchi@du.ac.ir}
\address{School of Physics, Damghan University, P.O. Box 36716-41167, Damghan, Iran}
\address{School of Physics, Institute for Research in Fundamental Sciences (IPM), 19395-5531, Tehran, Iran}

\date{\today}
\vspace{1cm}
\newbox\absbox
\begin{abstract}
We investigate transport properties through nano-ribbons of thin topological insulators irradiated by high frequency light with circular polarization. By using high frequency regime, a coherent and quantized transport through the nano-ribbon is guaranteed and then Lanadauer formalism is applicable. It is demonstrated that the pseudo-spin edge modes inside the band gap can host transmission through this nano-junction which their localization on the top and bottom edges depend strongly on the light polarization. These edge modes persist even if we apply a source-drain bias. Based on this edge selectivity for the current, one can design a light-induced switch with an appropriate on/off ratio of the current which is composed of two scattering regions with opposite light polarization. The local current on each bond shows how the current is passing through the edges and jumps into the opposite edge. Furthermore, some other nano-junctions are proposed as electronic switches which are designed based on the mass term engineering of the scattering region by means of perpendicular magnetization induced by magnetic doping and also structure inversion asymmetry applied on the scattering region.
\end{abstract}
\maketitle

\section{Introduction}
To achieve the best control on electronic devices, it seems that considering two dimensional (2D) materials is necessary and unavoidable. Although many 2D semiconductors have been investigated for this purpose, no one could overcome all challenges in practice. The most important challenge for new 2D materials is a significant mobility compared to silicon. Although Graphene as an intrinsically thin material wins the mobility competition, it has no band gap to be used in the field effect transistors (FETs) and any try to induce a band gap reduces its mobility drastically\cite{Ouyang}. On the other hand, sensitivity of the band structure and mobility of electrons to any edge roughness, imperfection and edge passivation, decommissions graphene nanoribbon as a good candidate for FET\cite{Fischetti}. Transition metal dichalcogenides, in spite of their advantages, have low mobility and high level of defects \cite{McDonnell}. Phosphorene and also Silicene have also very low mobility in practice\cite{Vandenberghe}. 

Among all materials used in conventional FETs, topological insulators (TIs) represent the best promising characteristic for possible applications in electronic devices \cite{Fischetti2017}. TIs are quantum materials which behave as insulator in the bulk while depending on their dimension, there are conducting states on their surface or edges. The surface (edge) states are protected by time reversal symmetry (TRS) which persist even at high level of imperfections such as vacancies, doping and impurities. So electronic devices fabricated based on TIs operate in low power with high performance\cite{Fischetti2017}. Many investigations have been implemented to fabricate FETs from TIs since their discovery\cite{Ezawa,Chang,Fischetti2017}. Among all approaches for designing a switch in TIs, making a control on phase transitions, especially by inducing a gap in the surface (edge) states, is a straightforward way in practice. 

In this work, a FET is designed based on the quantum anomalous Hall insulator (QAHI) phase. The QAHI phase was initially observed in TI thin films which are doped with magnetic impurities, such as Cr \cite{113137201,nature10731} and V \cite{qahv} doping in $(Bi,Sb)_2Te_3$. Materials such as $Bi_2Se_3$ \cite{wang2012}, $Sb_2Te_3$ \cite{Kong}, $Bi_2Te_3$ \cite{Pramanik} and $HgTe$ \cite{hgte} have a strong spin-orbit coupling which exhibit them as 3D topological insulators. The gapless surface states with Dirac cone dispersion on their surfaces is the characteristic of such materials. These two Dirac cones overlap with each other in a thin version of the above materials giving rise to an effective two dimensional insulator. 


Recent improvement in mid-infrared lasers, opened the opportunity for engineering the electronic band structure and also inducing topological phases in trivial materials illuminated by a polarized light. Inducing topological phases and even switching between them have been shown to emerge in HgTe quantum wells \cite{shelykh}, cold atoms \cite{Reichl, Jotzu} conventional insulators \cite{2011Lindner,2013Gomez,katsnelson,ando}and semi-metallic materials \cite{oka,kitagawa,driven,fweyl,Gonzalo,Wang} by application of time periodic driving fields. 
Moreover, the Floquet-Bloch states\cite{2013Gomez,cold2} have been experimentally observed in irradiated topological insulator $Bi_2Se_3$ illuminated by circularly polarized light \cite{obs,Mahmood}. 

In particular, in high frequency regime in which the driven frequency is much larger than the band width, one can use an effective static Hamiltonian which is composed of series expansion in inverse powers of frequency\cite{2013Gomez, ezawasilicene}. One of the advantages of off-resonant regime, is the population of the Floquet states which are occupied as the equilibrium systems\cite{Amitra}. Furthermore, in off-resonant regime, heating rate is low which is in favorite for an electronic device to operate properly\cite{Rudner_nature2020}.


In this paper, we investigate the transport properties of a nano-junction of thin topological insulator illuminated by circularly polarized light in off-resonant regime by means of Lanadauer formalism. In this work, a light-induced switch is proposed which is composed of two illuminated scattering region with distinct circularly polarization of light. Here, there is no need to the gap closing while switching is implemented by inducing two distinct pseudo-spin polarized QAHI phases only by means of a change in the polarization of light without application of any other perturbation. However, phase transition between different phases can be also occurred by means of other parameters such as magnetization, structural inversion asymmetry (SIA), thickness of the thin film and also light intensity. The details of switching phenomena and also edge currents are clearly explained by the current distribution through the TI nano-ribbon. It is worth to note that, the edge states decay exponentially into the bulk region and penetration depth is inversely proportional to the band gap. To avoid hybridization of the edge states, the nano-ribbon width is considered to be much wider than the penetration depth.

This paper is organized as follows: In Sec. \ref{S2}, we present low-energy Hamiltonian of two-dimensional topological insulator in the dark mode and by means of the high frequency expansion formalism \cite{Bukov,eckart,dabiri1}, an effective static Hamiltonian is presented for the driven system. The effective Hamiltonian in the absence and presence of SIA potential is extracted in Secs.\ref{S2_a} and \ref{S2_b} respectively\cite{dabiri2}. In Sec. \ref{S3}, transport properties of a TI nano-ribbon is presented by Landauer formalism. Finally, we submit the result and conclusion in Sec. \ref{S4} and \ref{S5} respectively.

\section{Dark and photo-assisted Hamiltonians}\label{S2}
Light-induced switch introduced in this paper is based on two-dimensional topological insulators designed on a thin film of $Bi_2Se_3$ and (Bi,Sb)$_2$ Te$_3$ family materials \cite{nature2009}. There are two Dirac cones on each surface which can be hybridized if the film is thinner than $5 nm$ \cite{nature584,prb81041307}. Tunneling between the upper and lower surfaces leads to a band gap opening in the spectrum of the surface edge modes such that the system is inverted to a 2D insulator. Interestingly, by means of magnetic impurities such as Ti, V, Cr and Fe, and also structure inversion asymmetry, the topological invariant of the system is changed and the edge modes may re-appear inside the gap. It should be noted that irradiation of light at high-frequency regime leads to a rich feature of phase diagram in which without application of an external magnetic field, by tuning light parameters such as amplitude and polarization and also induced magnetization, one can engineer different topological phases such as quantum pseudo-spin Hall insulator or quantum anomalous Hall insulator\cite{dabiri1,dabiri2}.
The low energy effective Hamiltonian for the {\it dark mode} around $\Gamma$ point is written \cite{effective,ryu} as the following; 
\begin{equation}
\begin{aligned}
\centering
&H_{\text{dark}}(\textbf{k})=
\hbar {v}_f \tau_z\otimes (k_y \sigma _x -k_x \sigma _y) 
+\Delta (\textbf{k})\tau_x\otimes \sigma _0 
\\&+V_{SIA}\tau_z\otimes \sigma _0 
+ M_z \tau_0\otimes\sigma_z 
\end{aligned}
\label{eq:firsthamil}
\end{equation}
Without loss of generality, we have neglected particle-hole symmetry term in this Hamiltonian. The basis set in which the above Hamiltonian is written in is represented as
$|t,\uparrow \rangle ,|t,\downarrow \rangle ,|b,\uparrow \rangle ,|b,\downarrow \rangle$, where $t (d)$ refers to the top (bottom) surface states and $\uparrow (\downarrow)$ displays the up (down) spin state. The matrices $\tau_i (\sigma_i)$ in the Hamiltonian are Pauli matrices in the surface space (spin space). The surface state as a Dirac cone spectrum with Fermi velocity of $v_f$ appears in the first term.
The second term comes from tunneling between these Dirac cones of surfaces in which the experimental fittings result in the following k-dependent tunneling $\Delta(\textbf{k})=\Delta_0+\Delta_1 k^2$ for the thin film of Bi$_2$ Se$_3$ and [(Bi,Sb)$_2$ Te$_3$] family\cite{nature584,prb81041307}. The parameters $\Delta_0$ and $\Delta_1 $ which depend on the thickness of TI thin film are determined by the experimental data. The third term is related to the structure inversion asymmetry $V_{SIA}$ could be generated by using either of two factors: the perpendicularly applied electric field or substrate effects\cite{nature584}. Finally, the last term expresses the effect of induced exchange field arising from doped magnetic impurities in Bi$_2$Se$_3$ and [(Bi,Sb)$_2$ Te$_3$] family\cite{Magnetic}. In this work, we consider system parameters as $\Delta_0=35$~meV , $\Delta_1=-10$~eV\AA$^2$ , $v_f=4.48 \times 10^{5}$~m/s, and $\hbar \Omega=1$~eV as driven frequency which is much larger than the band width and stands for an off-resonant regime in which the central Floquet band is far from other replicas.

A change in the basis set by means of a unitary transformation does not change the eigenvalues of the system. Let us write the new basis set by using the bonding and antibonding states as$\vert \psi_b,\uparrow\downarrow\rangle=\left( \vert t,\uparrow\downarrow\rangle+\vert b,\uparrow\downarrow\rangle \right)/ \sqrt{2}$ and
$\vert \psi_{ab},\uparrow\downarrow\rangle=\left( \vert t,\uparrow\downarrow\rangle-\vert b,\uparrow\downarrow\rangle \right)/\sqrt{2}$. Now one can derive the new form of the Hamiltonianin in the basis of $|\psi_b,\uparrow \rangle ,|\psi_{ab},\downarrow \rangle ,|\psi_b,\downarrow \rangle ,|\psi_{ab},\uparrow \rangle$ as 
\begin{equation}
\begin{aligned}
H^{\prime}_{\text{dark}}(\textbf{k})=&\text{$\hbar $v}_f \left(k_y \tilde{\tau}_0\tilde{\sigma} _x- k_x\tilde{\tau}_z \tilde{\sigma} _y \right) \\
&+\Big(\Delta(\textbf{k}) \tilde{\tau}_0+ M_z \tilde{\tau}_z \Big)\tilde{\sigma} _z+V_{SIA}\tilde{\tau}_x \tilde{\sigma}_x
\end{aligned}
\label{eq:total} 
\end{equation}

in which $\tilde{\tau}_i$ and $\tilde{\sigma}_i$ $(i=0,x,y,z)$ are Pauli matrices in the bonding and anti-bonding and also spin Hilbert space . 
\begin{figure*}
\centering
\includegraphics[width=0.6\linewidth]{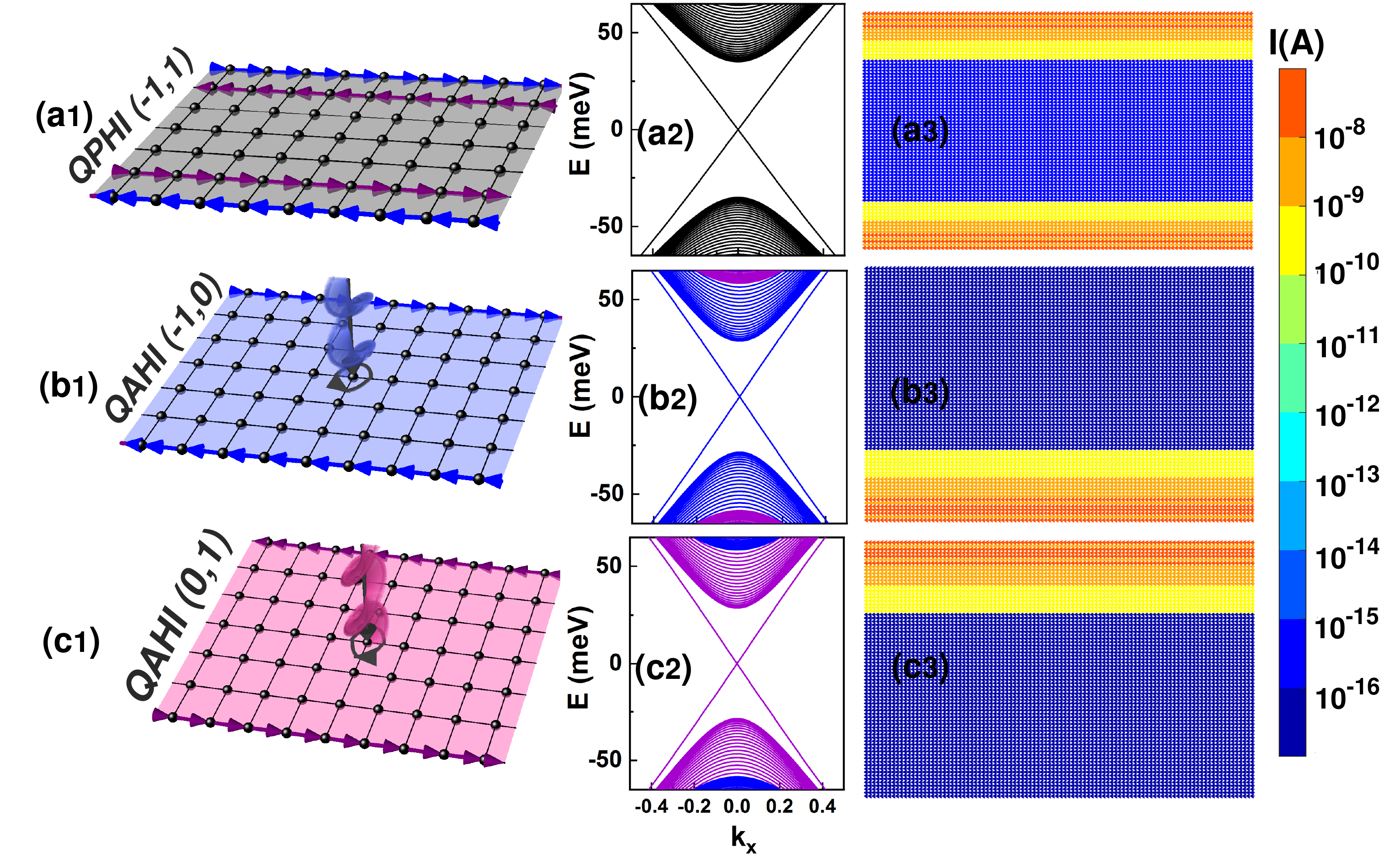}\includegraphics[width=0.4\linewidth]{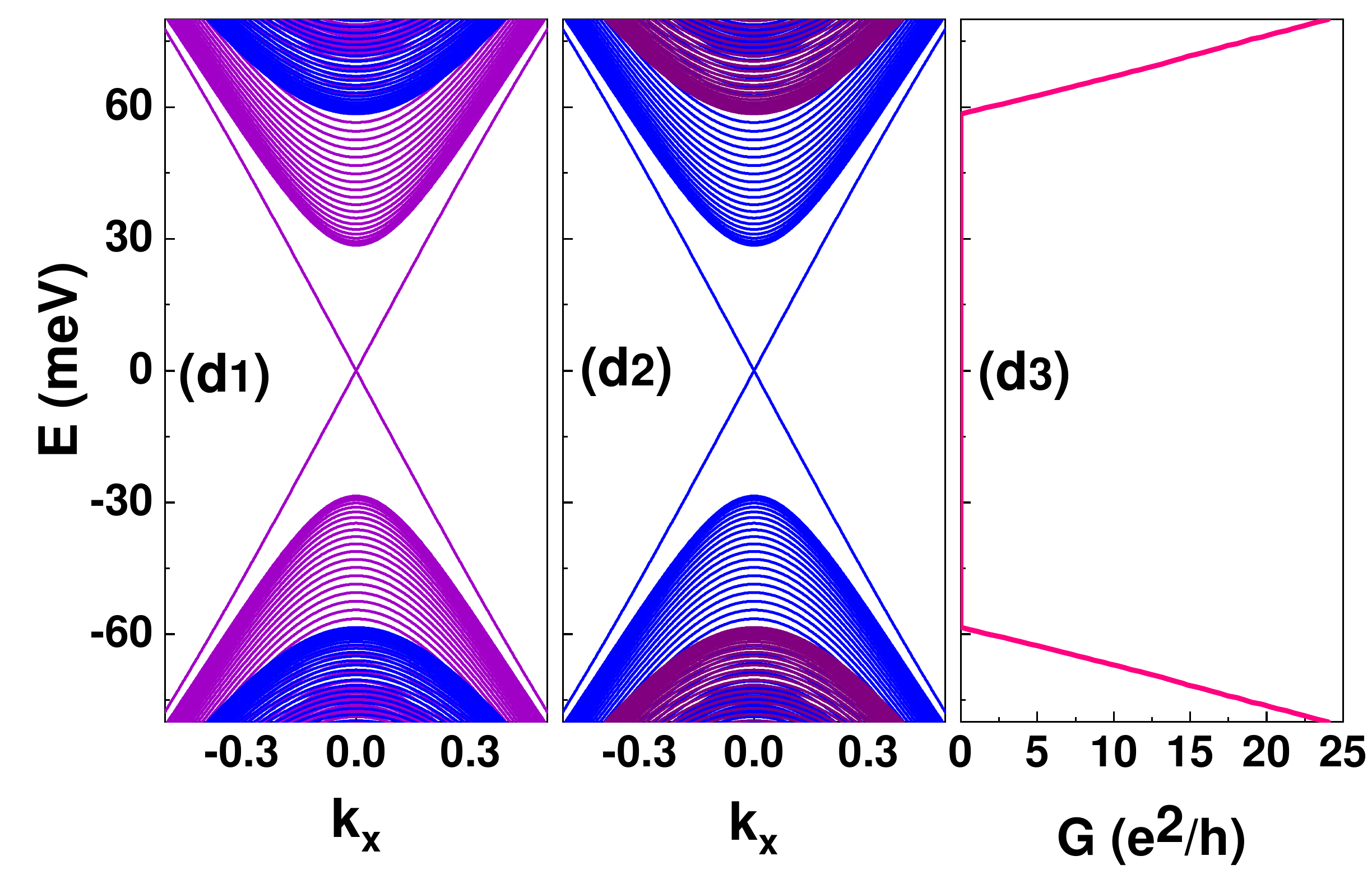}
\caption{(a1) A nano-ribbon version of thin TI with aforementioned parameters which lies in QPHI phase. In low energies, two helical edge modes which are pseudo-spin polarized are shown by the blue (pseudo-spin +) and purple (pseudo-spin -) arrows. (a2) band structure of the dark nano-junction in which black bands indicate degeneracy of two pseudo-spin polarized bands. (b1 and c1) the system illuminated by ( RCP and LCP) light so that a phase transition from QPHI to QAH phase occurs. In low energies, only one pseudo-spin polarized edge mode ($+$ and $-$) is hosted which is shown by the (blue and purple) arrows. (b2 and c2) band structure of a thin topological insulator illuminated by (RCP and LCP) light in which blue (purple) bands show pseudo-spin + (-)states. (a3, b3, c3) Current distribution of the scattering region sandwiched between two dark electrodes under application of source-drain potential confirms proposed distributions based on topological phases in each case. (d3) Conductance of the nano-junction composed of a step mass-term structure made by topological insulator nano-ribbons irradiated by LCP (left portion) and RCP (right portion) light. (d1, d2) the band structure of the left and right portion of this step mass-term structure.}
\label{fig1}
\end{figure*}
\subsection{Perpendicular magnetization at zero SIA Potential}\label{S2_a}
In this new form of Hamiltonian, it is clear that if $V_{SIA}=0$, the above effective Hamiltonian \ref{eq:total} would be blocked with the diagonal matrices labeled by the pseudo-spin index ($\alpha=\pm$) as the following\cite{universality,geometrical,PRL2013_zhang}

\begin{equation}
\centering
h^{dark}_{\alpha }(\textbf{k})=\text{$\hbar $v}_f \left(k_y \tilde{\sigma} _x- \alpha k_x \tilde{\sigma} _y\right)+\Big(\Delta(\textbf{k})+\alpha M_z \Big)\tilde{\sigma }_z
\label{eq:hamilchiral} 
\end{equation}

Illumination of light with circular polarization on a TI thin film is simulated by the following time-periodic vector potential; $\overrightarrow{A}(t)=A_0(\sin(\Omega t),\cos(\Omega t))$ where $|\Omega|=2 \pi /T$ is the frequency of the drive. The sign of $\Omega$ refers to the left or right-handed polarization. Thanks to the Peierls substitution which is a change in the wave vector induced by the vector potential as the following: $k_i \longrightarrow k_i+\frac{e A_i}{\hbar}$. It is straightforward to substitute this wave vector in Hamiltonian \ref{eq:firsthamil} giving rise to a time-periodic Hamiltonian. This type of Hamiltonian is studied by means of a powerful theorem named as Floquet theory. In this theory, one can define a Floquet Hamiltonian which describes system evolution in stroboscopic times which is multiplier of the period of vector potential $T$. The Floquet-Schroedinger equation is represented as
 \begin{equation}
\centering
[H_F(t)-i\hbar\frac{\partial}{\partial t}] |\phi_{\alpha}(t)\rangle=\varepsilon_\alpha |\phi_{\alpha}(t)\rangle
\end{equation}
where $|\phi_{\alpha}(t)\rangle$ is the Floquet state which obeys the periodicity of the vector potential as well as $H(t)$. Here, the quasi-energy of this Hamiltonian, $\varepsilon_\alpha$, is formed in a band spectrum which contains central Floquet bands accompanying by infinite sidebands. The Fourier transformation of the above equation is simply inverted to the following eigenvalue equation;
 \begin{equation}
\centering
(\varepsilon_\alpha +m \hbar \Omega)|\phi^{m}_\alpha\rangle=\sum_{m'}H_F^{(m-m')}|\phi^{m'}_\alpha\rangle
\label{inf}
\end{equation}

where the Fourier coefficient of Floquet Hamiltonian is defined as $H_F^{ m}=1/T \int_{0}^{T} H_F(t) e^{ im |\Omega| t} dt$. Fortunately, if the energy of incident photons is higher than any characteristic energy of the system (here the band width), and also at low intensity of irradiations, it is straightforward to find a series expansion for the Floquet Hamiltonian in terms of the inverse frequency $1/\Omega$ as the following photon-dressed Hamiltonian\cite{Bukov,eckart,bw},  

\begin{equation}
H_{eff.}=H_F^0+(\hbar \Omega)^{-1}[H_F^{-1},H_F^{+1}]+\mathcal{O}(\frac{1}{(\hbar\Omega)^2})
\label{one_photon_H}
\end{equation}
 where it is proved that $H^{i}=0$ (for $i\ne 0,\pm1$). Looking at the above effective Hamiltonian deduces that one-photon process is only assisted in transmission phenomena. Let us look at the effective Floquet Hamiltonian if $V_{SIA}=0$ in this regime. In this case, as we mentioned before, the dark Hamiltonian is pseudo-spin polarized even if perpendicular magnetization is applied on the film. After Peierls substitution in the dark pseudo-spin polarized Hamiltonian, one can simply drive the effective Hamiltonian as

\begin{equation}
h_\alpha=\hbar \eta_\alpha v_f ( k_y \tilde{\sigma}_x-\alpha k_x \tilde{\sigma}_y)+[\Delta'({\bf k})+\alpha(M_z+m_{\Omega})]\tilde{\sigma}_z
\label{Photo_Hamiltonian}
\end{equation}
 where

\begin{equation}
\begin{aligned}
&\Delta'(\textbf{k})=\Delta_0+\mathcal{A}^2\Delta_1+k^2\Delta_1 , ~~  A^\prime=\frac{ \mathcal{A}^2 }{\hbar \Omega} \\
&\eta_\alpha=1- 2\alpha A^\prime \Delta_1 , ~~ m_{\Omega}=\hbar^2v_f^2A^\prime
\end{aligned}
\end{equation}

In this formula, the scaled intensity is defined as $\mathcal{A}=eA_0/\hbar$. Regarding to the sign of mass term, the pseudo-spin Chern number is given by $\mathcal{C}_\alpha=\alpha/ 2 \big (\text{sgn}\Delta_1 -\text{sgn} (\Delta_0+\mathcal{A}^2\Delta_1+\alpha m) \big)$ where sgn refers to the sign function and $m=M_z+m_{\Omega}$. $m_{\Omega}$ plays the role of the mass term induced by the illuminated light. The total Chern number is written as $\mathcal{C}=\mathcal{C}_+ + \mathcal{C}_-$. The phase diagram of this Hamiltonian has been investigated in Ref.[\cite{dabiri1}] in detail. Depending on the sign of $\Delta_0 \times \Delta_1$ and light parameters such as the intensity, polarization of light and also system parameters such as magnetization $M_z$, there are two types of phase diagram in which there are regions with quantum anomalous Hall insulator, normal insulator or quantum pseudo-spin Hall insulator phases\cite{dabiri1}. The other interesting phenomena is {\it anisotropic helical edge} states appeared in the coefficient $\eta_\alpha$. In fact, the Fermi velocity can be different for each pseudo-spin such that $\delta \eta=\eta_--\eta_+=4 A^\prime \Delta_1$ depends on the light intensity, frequency and also hopping parameters between two surfaces.       

\subsection{In presence of SIA potential}\label{S2_b} 
 In the presence of SIA potential, pseudo-spin polarization fails and off-diagonal terms in Hamiltonian (\ref{eq:total}) is inverted to non-zero values\cite{sabzeh}. Therefore, we begin with the dark Hamiltonian indicated in Eq. \ref{eq:firsthamil}. After Peierls substitution and at high frequency regime in low intensity of the driven fields, the following Hamiltonian is deduced,
\begin{equation}
\begin{aligned}
H=& (1-2A'\Delta_1 \tilde{\tau}_z) \hbar v_f ( k_y \tilde{\sigma}_x- k_x\tilde{\tau}_z \tilde{\sigma}_y) \\
& +[\Delta'({\bf k})+\tilde{\tau}_z m]\tilde{\sigma}_z +V_{SIA}\tilde{\tau}_x\tilde{\sigma}_x
\label{Photo_Hamiltonian2}
\end{aligned}
\end{equation} 

The gap closing conditions for the above Hamiltonian demonstrate that after a critical SIA potential there would be a phase transitions from QAHI and also QPHI to NI depending on the initial phase at zero SIA potential. A detailed study is found in Ref.[\cite{dabiri2}].  
\begin{figure*}
\centering
\includegraphics[width=0.6 \textwidth]{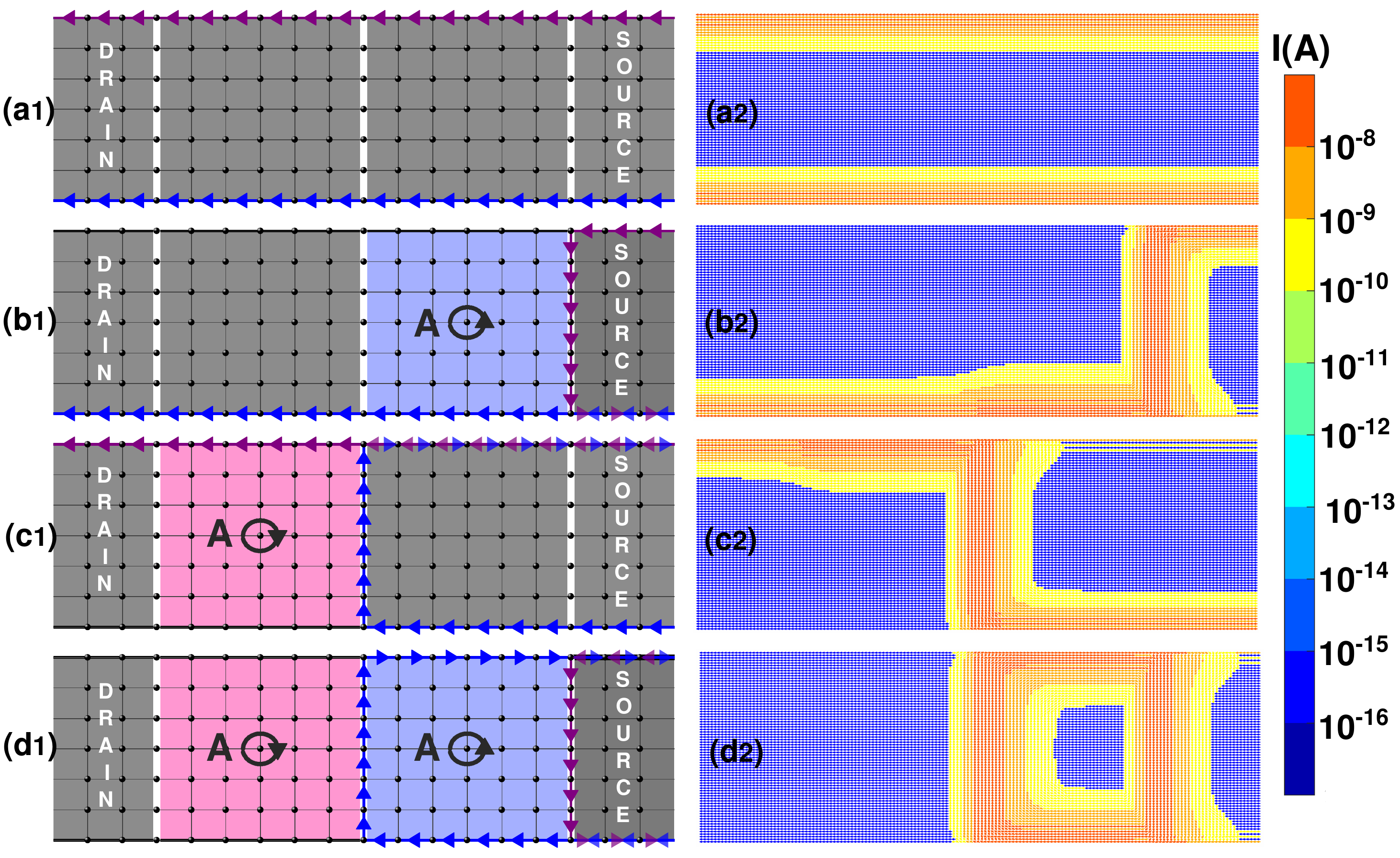} \includegraphics[width=0.35\textwidth]{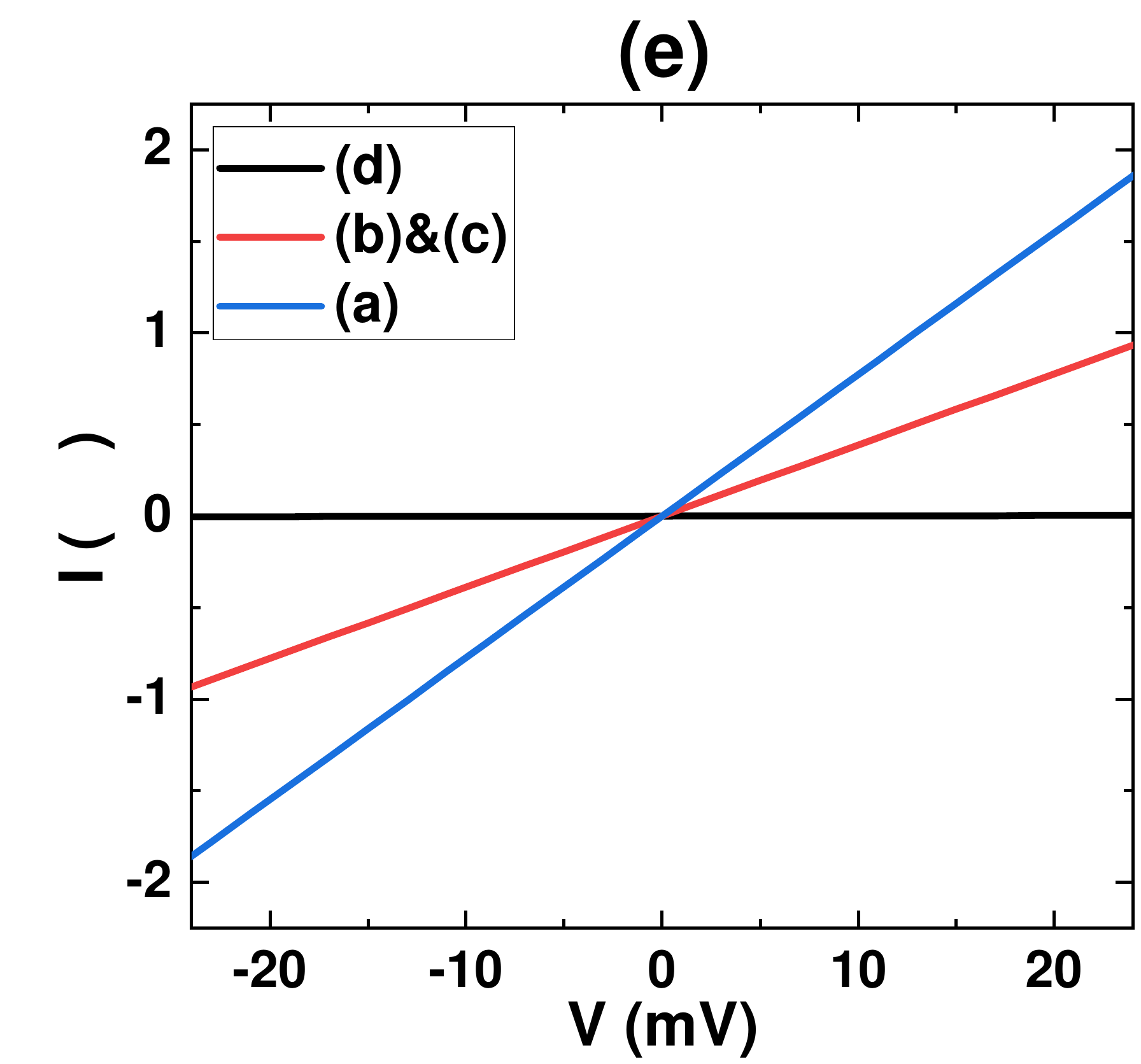}
\caption{Schematic view of four mass-term structure as a transistor with a switching operation induced by light polarization. The color and pigmentation value of vectors are the indicator of the pseudo-spin state and the relative magnitude of current. (a1) In the dark scattering region, both pseudo-spins flow through the transport direction. (b1 and (c1)) Irradiation of RCP (LCP) light on half portion of the scattering region which selects pseudo-spin $\alpha_z=+1$ ($\alpha_z=-1$) to tunnel through the region and filters the other one. (d1)Transport blocking arising from the alignment of opposite pseudo-spin states as a result of illumination on half portion of the scattering region with RCP light and the other half with LCP light, simultaneously. (a2, b2, c2 and d2) Real current distributions which consist of the scattering region and part of electrodes in which the color of vectors show only the real magnitude of current denoted by a color bar in the right side. (e) I-V characteristic curve. The blue (red) lines show {\it on-}state of the switch which are related to transmission of two (one) pseudo-spin states. The black line shows the {\it off-}state of the switch.}
\label{fig2}
\end{figure*}

\section{Tight-binding and Landauer Formalism} \label{S3}
To investigate transport properties of the edge modes of topological insulators by using Landauer formalism, it is convenient to cut the film into a nano-ribbon shape. To do this, let us discretize the Hamiltonian presented in Eq. \ref{eq:total} on a square lattice,
\begin{equation}
\label{TB_Hamiltonian}
H=\sum_{\bf r}^{} {\bf c}_{\bf r}^\dagger T_0{\bf c}_{\bf r}+
{\bf c}^\dagger_{\bf r+\hat{x}} T_x{\bf c}_{\bf r}+{\bf c}^\dagger_{\bf r+\hat{y}} T_y{\bf c}_{\bf r}+h.c.
\end{equation} 
where  ${\bf c}_{\bf r}^\dagger$ and ${\bf c}_{\bf r}$ are the creation and annihilation operators defined on the site $\bf{r}$, 
and 
\begin{equation}
\begin{aligned} 
T_0=&\left(\Delta_0^\prime+\frac{4\Delta_1}{a^2}\right)\tilde{\tau}_0\tilde{\sigma}_z+m\tilde{\tau}_z\tilde{\sigma}_z+V_{SIA}\tilde{\tau}_x\tilde{\sigma}_x,\\
T_x=&-\frac{\Delta_1}{a^2}\tilde{\tau}_0\tilde{\sigma}_z+i\frac{\hbar v_f}{2a}(\tilde{\tau}_z\tilde{\sigma}_y-2A'\Delta_1\tilde{\tau}_0\tilde{\sigma}_y)\\
T_y=&-\frac{\Delta_1}{a^2}\tilde{\tau}_0\tilde{\sigma}_z-i\frac{\hbar v_f}{2a}(\tilde{\tau}_0\tilde{\sigma}_x-2A'\Delta_1\tilde{\tau}_z\tilde{\sigma}_x)\\
\end{aligned}
\end{equation}
are 2$\times$2 onsite-energy ($T_0$) and the hopping-energy matrices ( $T_x$ , $T_y$ ) along the $x$ and $y$  directions, respectively. Here, $\alpha_z$ is pseudo-spin index. $y=ja$ is the site position along $\hat{y}$ direction.
The current flowing through the scattering region can be expressed by Landauer formula presented as the following:
\begin{equation}
\begin{aligned}
I(V_{SD})=\\
\dfrac{e}{\hbar}\int_{\mu_{L}}^{\mu_{R}}T(E,V_{SD})( f(E-\mu_{L})-f(E-\mu_{R}))dE&  
\end{aligned}
\end{equation}
Where $f(E,E_{f})$ is the Fermi-Dirac distribution function and $\mu_{L} (\mu_{R})$ is the chemical potential of the left (right) electrode. $T(E,V_{SD})$ is the transmission coefficient which is obtained by $T=Tr[\Gamma_{L}G_{C}\Gamma_{R}G_{C}^{\dagger}]$, where $G_{C}(G_{C}^{\dagger})$ is the retarded (advanced) Green's function of the scattering region and $\Gamma_{L}(\Gamma_{R})$ is the coupling function of the scattering region and left (right) electrode.

In the presence of the source-drain bias, the local charge current from the site position ${\bf r}$ to the other site ${\bf r} + {\bf r}_{0}$ is calculated by:
\begin{equation}
\begin{aligned} 
J_{{\bf r},{\bf r}+{\bf r}_{0}}=\\
\dfrac{e}{h}\int_{-\infty}^{\infty}dE[H_{{\bf r},{\bf r}+{\bf r}_{0}}G_{{\bf r}+{\bf r}_{0},{\bf r}}^{<}(E)-H_{{\bf r}+{\bf r}_{0},{\bf r}}G_{{\bf r},{\bf r}+{\bf r}_{0}}^{<}(E)]&
\end{aligned}
\end{equation}
in which the lesser Green's function is calculated by the Keldysh equation \cite{keldysh}.
\begin{equation}
G^{<}(E)=G^{r}(E)\Sigma^{<}G^{a}
\end{equation}
where the lesser self-energy is defined in terms of the left and right self-energies induced by the electrodes,
\begin{equation}
\Sigma^{<}=\sum_{L(R)}if_{L(R)}(E)\Gamma_{L(R)}(E).
\end{equation}

   Before presenting the results, let us notice that in off-resonant regime, there is no photon-assisted transitions between the central bands and Floquet side-bands. Therefore, in this regime, transport is a coherent phenomena, while in on-resonant regime, transport is non-coherent. Indeed, at low frequency regime, by absorbing or emitting of "n" photons, energy of incident electron entering from the non-driven electrode will be changed in the scattering region which is called "photon-assisted" tunneling. In this case, regarding to imperfect energy matching between non-driven states in the dark electrodes and the side-band states (Photon-assisted states), quantization of Hall conductance at zero energy arising from topological states fails\cite{Rudner_nature2020}. To protect the edge modes during the tunneling, an energy filter is used in the electrodes to suppress photon-assisted tunneling. As presented before, in high frequency regime, Hamiltonian in the irradiated scattering region is approximated with an effective static-like Hamiltonian which describes Floquet central bands. So we expect to trace these edge modes during transport measurements.
    
    The light-induced switch that we have considered, is a nano-ribbon version of thin topological insulator which is irradiated by high-frequency laser field giving rise to a control on the current passing through the mesoscopic system. As shown in Fig.~\ref{fig1}, illumination is limited to the scattering region and two electrodes are in the dark states. Moreover, the scattering region, can be essentially doped by magnetic impurities which causes to change the gap and also topological phases of the scattering region. Moreover, a structural inversion asymmetry (SIA) can be applied on the scattering region leading to an accessible electrical switch.

\begin{figure*}
\centering
\includegraphics[width=0.7 \textwidth]{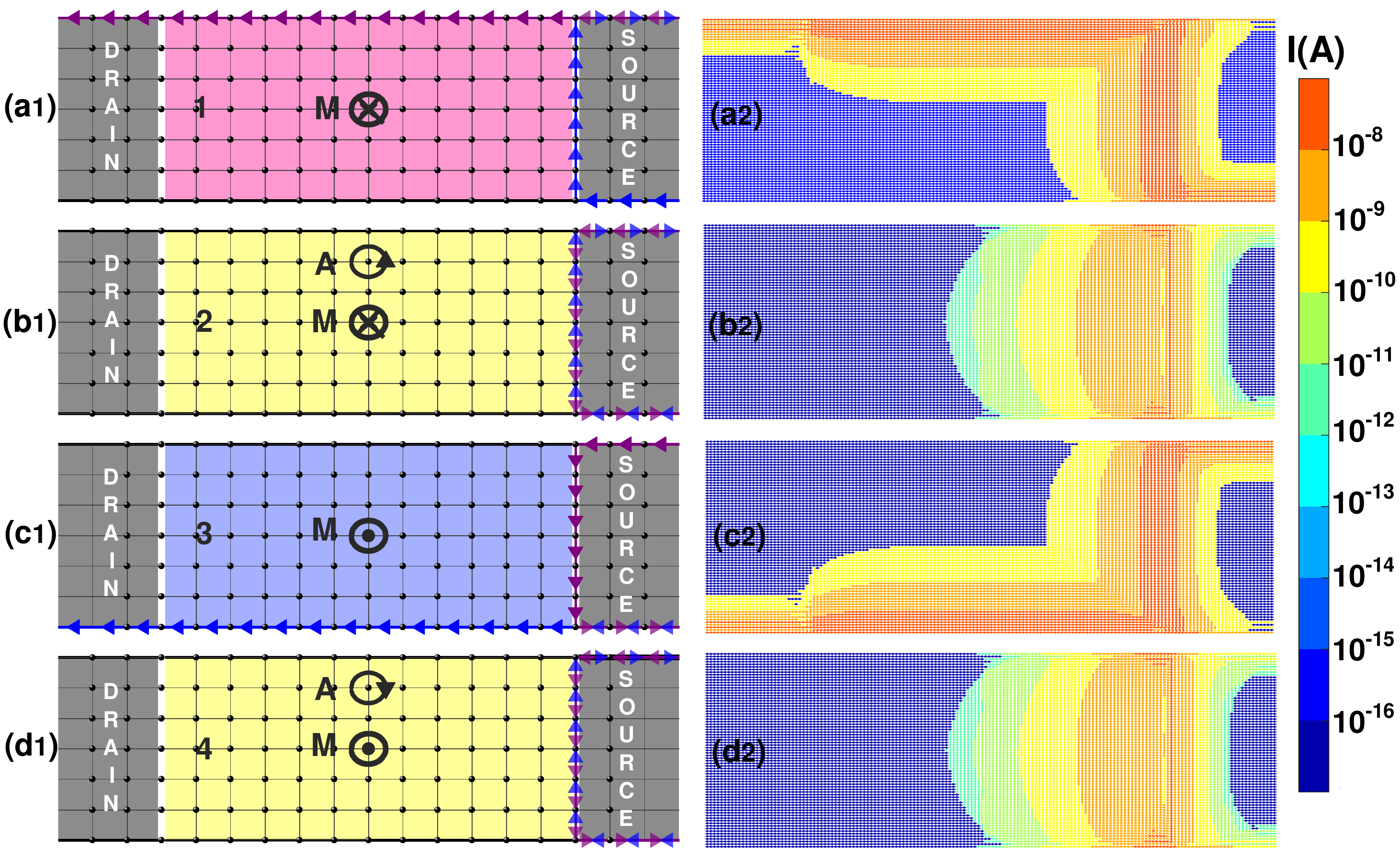}\includegraphics[width=0.3\linewidth]{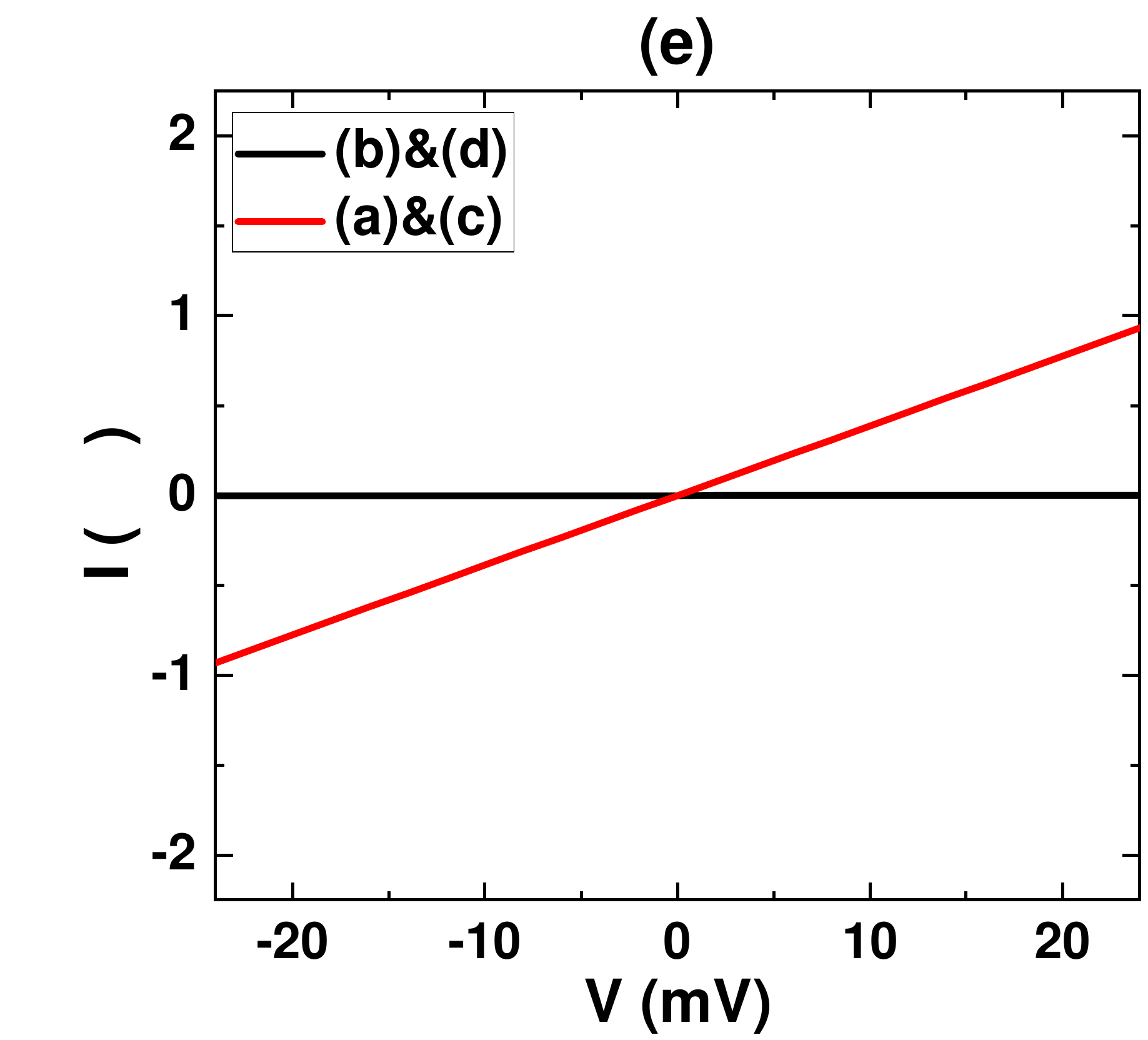}  
\caption{A magnetically doped transistor switched by light polarization. (a1 and c1) Edge selectivity of current by the direction of magnetization arising from magnetic doping in the dark mode. (b1 and d1) applying both magnetization and circularly polarized light at the same time converts the scattering region into NI phase and leads to the {\it off-}state of the switch. Current distributions of schematic drawing presented in a1, b1, c1 consist of the scattering region and electrodes show (a2 (c2)) flowing $\alpha_z=+1$ ($\alpha_z=-1$) pseudo-spin states through the top (bottom) edge. (b2 and d2) blocking both pseudo-spin states in the scattering region which is in NI phase. (e) I-V characteristic curve of the switch. The red curve shows {\it on-}state of the switch which is related to flow of one pseudo-spin state and the black curve shows {\it off-}state of the switch.}
\label{fig3}
\end{figure*}

\section{Results}\label{S4}
Before turning on the irradiation, at first, let us investigate switching phenomena in the absence of SIA potential and also lack of magnetic impurities. Regarding to the phase diagram shown in Ref.\cite{dabiri1}, with a choice of $\Delta_0 \Delta_1 <0$, all three portions of the nano-junction is in the QPHI phase. At low energies, two pseudo-spin polarized edge states are flowing through the nano-junction helically. As it was noticed before in Eq.~\ref{Photo_Hamiltonian}, depending on the light intensity and frequency of the pump, anisotropic helical states essentially emerge which have different Fermi velocities. However, by our considered parameters, this difference is small enough to affect the band spectrum and also transport phenomenon.

A schematically view of the dark nano-junction is drawn in Fig.~\ref{fig1}(a1) which shows the $+$ (blue) and $-$ (purple) pseudo-spin polarized edge currents. Moreover, the band spectrum of this nano-ribbon confirms emergence of these pseudo-spin edge modes inside the band gap which are degenerated. 

In this stage, scattering region is irradiated by the left and right-handed circularly polarized (LCP and RCP) light as shown in Figs.~\ref{fig1} (b1, c1). So the mass term is varying and the topological phase of the system changes to QAHI. As shown in Figs.~\ref{fig1} b1 and c1, polarization of the light opens a selectivity for choosing special kind of pseudo-spin edge mode to flow through the nano-ribbon.

By applying source-drain bias, one of the edge currents is intensified along the transport direction, while the edge current flowing through the opposite direction is removed. In this work, the right electrode operates as source. Current distributions which are calculated by Landauer formalism are shown in Figs.~\ref{fig1} (a3, b3, c3). These distributions confirm the existence of the edge current as well as their helical property. In a QPH phase or in the case of dark nano-ribbon, Fig.~\ref{fig1} (a1) , for example, because of source-drain external bias, those pseudo-spin states which are allowed along the transport direction contribute helically in the current. 

At zero SIA potential and also zero in-plane magnetization, Hamiltonian of Eq.~\ref{Photo_Hamiltonian} is pseudo-spin polarized.  By using this property, a pseudo-spin selective tunneling occurs at a {\it mass-term step} nano-junction of two regions which are illuminated by RCP and LCP light. It is simply verified that in this case, because of alignment of the edge modes with opposite pseudo-spin states, a transport gap emerges in this nano-junction. Figs.~\ref{fig1} d shows the band spectrum of two regions illuminated by the RCP and LCP light and its resultant conductance through nano-junction which is calculated by non-equilibrium Green's function formalism.    

\begin{figure*}
\center
\includegraphics[width=0.7 \textwidth]{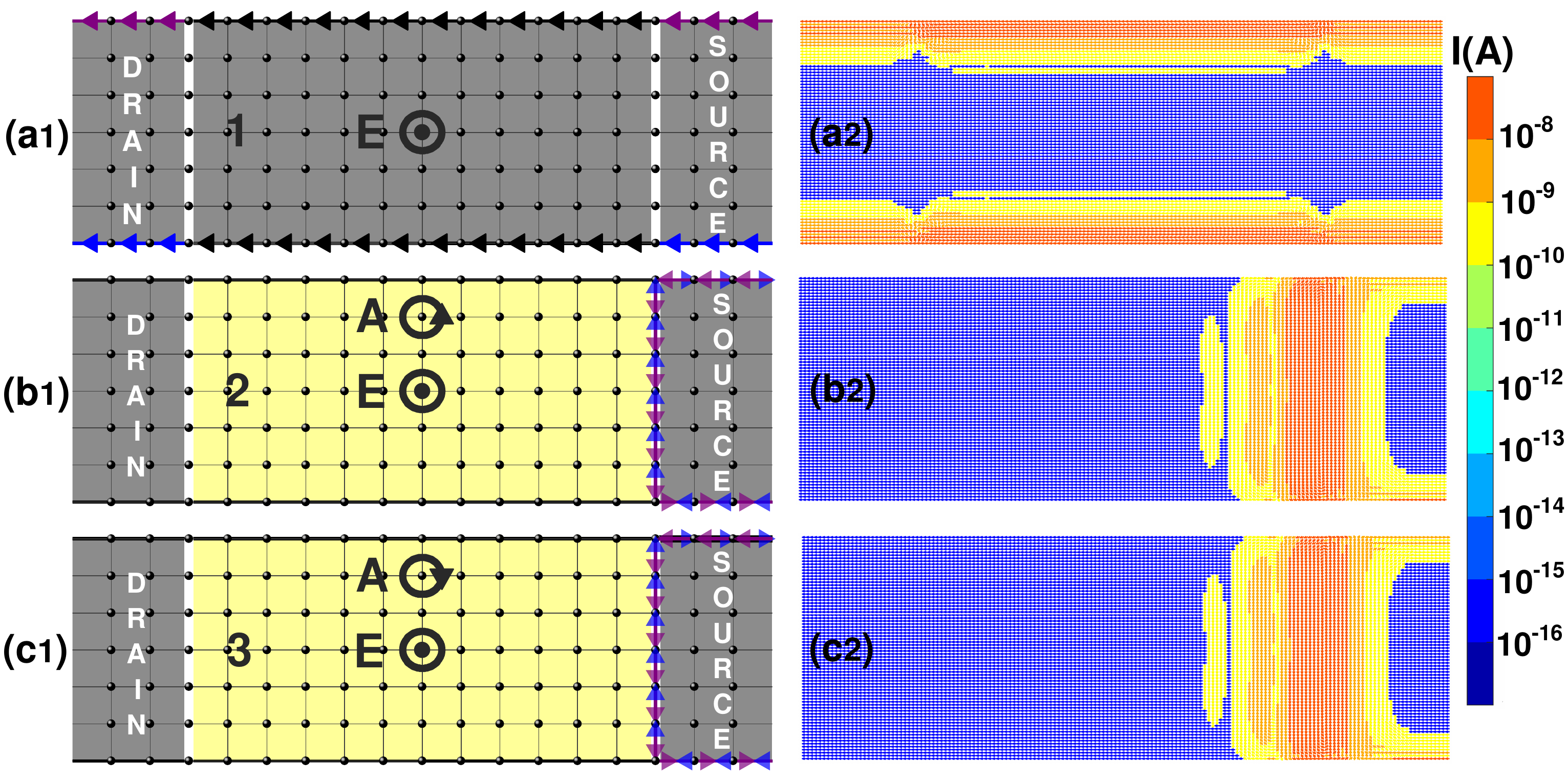}\includegraphics[width=0.3\linewidth]{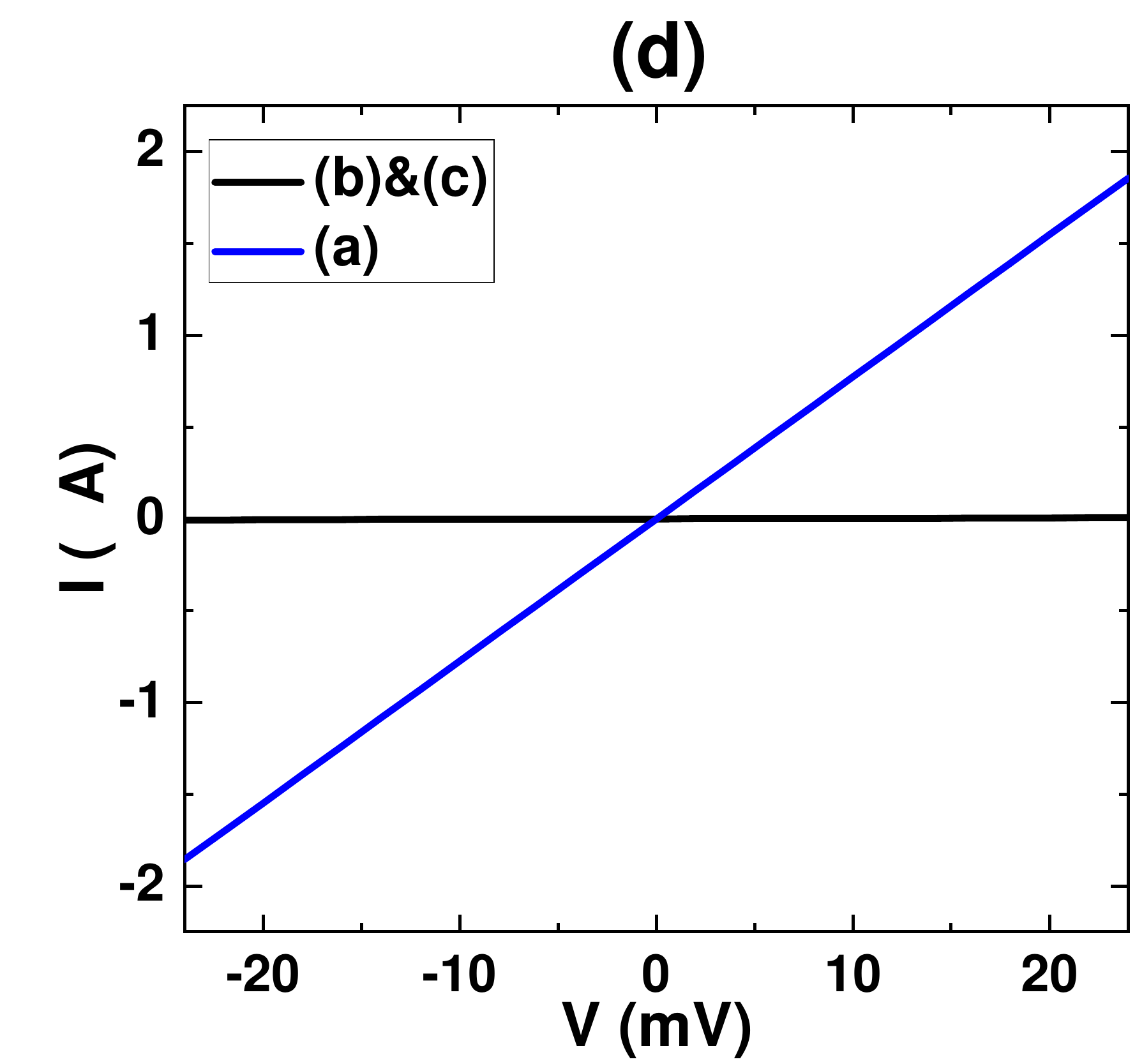}  
\caption{A field effect transistor switched by the light polarization. (a1) independent of the direction of perpendicular electric field, the scattering region lies in QPHI phase with mixed pseudo-spin states displaying by the black arrows. (b1 (c1)) A NI phase achieved by applying an electric field and light illumination on the scattering region at the same time. Pseudo-spin polarized states coming from the source electrode are reflected and do not flow through the scattering region. This is the {\it off-}state of the switch achieved by RCP (LCP) light. Local currents which show (a2) the mixing of pseudo-spin states arising from SIA. (b2 , c2) current blocking arising from NI phase as a result of applying both electric field and light irradiation. (d) I-V curve of switcher in which the blue curve is equivalent to the {\it on-}state when two pseudo spin states passing through the nano-junction. The black curve is equivalent to the {\it off-}state.}
\label{fig4}
\end{figure*}

 By taking this significant selective rule, a four mass-term structure is designed on thin topological insulator as a switcher induced by the light polarization. In the dark mode (QPHI phase), as far as source-drain bias is applied, the $+ (-)$ pseudo-spin edge mode shown schematically in Fig.~\ref{fig2} a1, flows through the transport direction from the down (up) edge. This analysis is verified by calculating current distribution represented in Fig.~\ref{fig2} a2 which is done by means of non-equilibrium Green's function formalism. The current is concentrated on the edges of the nano-ribbon which is consistent with schematic drawing in Fig.~\ref{fig2} a1. 
 
Now let us depart the scattering region in two parts illuminated separately. For the first, as shown in Fig.~\ref{fig2} b1, the right portion is illuminated by a RCP light while the left portion remains in the dark mode. Because of the voltage gradient, both pseudo-spin states are contributed in the edge currents; however, once the RCP light is illuminated on the right portion, the only pseudo-spin which is allowed to tunnel through the right portion would be $\alpha_z=+1$ flowing through the lower edge. So the pseudo-spin states with $\alpha_z=-1$ are filtered by the light illumination. Moreover, the current at the upper edge is blocked, so the polarization of circularly polarized light enables selecting on which edge, the current is permitted to flow. By using NEGF calculations, current distribution is depicted in Fig.~\ref{fig2} b2 confirms what we have proposed in Fig.~\ref{fig2} b1. At the lower edge but inside the source electrode, the edge current is nearly nullified that is explained by returning the edge current from the portion illuminated by RCP light.
 
 The same phenomenon happens provided that the left portion is illuminated by LCP light and as what is seen in Figs.~\ref{fig2} c1 and c2, the right portion is in the dark mode. In this turn, just the pseudo-spin $\alpha_z=-1$ is permitted to flow through the upper edge. Again the polarization of light selects the location of the edge current which can be used as a current splitter induced by the light polarization. The electrons with pseudo-spin $\alpha_z=+1$ are reflected from the junction created between the dark and illuminated region. Because of this reflected pseudo-spin polarized states, local current distribution in the right-top edge of the nano-ribbon shows very low current values. 
 
Finally, in Figs.~\ref{fig2} d1 and d2, we checked the situation in which both the left and right portions are illuminated by LCP and RCP light, respectively.  As a result, no current passes through the nano-junction and this electronic switch is in its off-current mode. Note that the local current distribution in the right portion of this nano-junction is only originated from $\alpha_z=+1$ states which are reflected from the junction between two illuminated regions with opposite polarity. Most current distribution is concentrated on the right portion which is irradiated by RCP light. If one look at the edge current inside the source electrode (the right side), the helical states with opposite pseudo-spins flow in opposite directions at each edge giving rise to low values of the current.

The current-voltage characteristic curves for all the above cases of the mass-term step junctions are summarized in Fig.~\ref{fig2} e. The I-V curve shows an Ohmic behavior in three different types. The current intensity at a fixed external bias, is twice in the dark mode than the current intensity measured in the configuration in which just one portion is illuminated by the polarized light depicted in Figs.~\ref{fig2} b, c. In the {\it off} state shown in Fig.~\ref{fig2} e with the black solid line, the output current is nullified as far as the voltage window of the external bias is lower than the system gap. The Fermi energy is fixed at zero energy. The {\it on/off} ratio at bias voltage $V=10 mV$ is in the order of $10^{2}$.

 In this stage, we investigate the effect of perpendicular magnetization and also applied electric field in combination with irradiation on the current passing through the nano-ribbon. First of all, we need to review different phases emerging for various parameter values. For the sake of completeness, we refer the reader to the phase diagrams drawn in Figs.~\ref{A1}, \ref{A2} in which the specific points $\alpha_1$ , ... , $\alpha_4$ or $\beta_1$ ... $\beta_3$ are marked, respectively.

For example, in a switcher represented in Fig.~\ref{fig3}, the mass term of the scattering region is controlled by a combination of the light parameters and magnetization. As seen in the phase diagram shown in Fig.~\ref{A1}, in the dark mode ($\alpha_1$ and $\alpha_3$), there are two QAHI phases with different chern numbers and consequently different edge channels for transport which are induced by magnetic doping. If the downward magnetization changes to the upward direction, the location of the chiral-polarized edge mode hops from the top to the bottom edge of the nano-ribbon. All these phases can be implemented in TI nano-junction. In Figs.~\ref{fig3} a1 and c1, the scattering region is doped by magnetic impurities in opposite directions. As it is clearly observed in Figs.~\ref{fig3} a2, c2, looking at the local current distribution which is calculated by NEGF formalism, demonstrates the localization of the edge transport channels in the top and bottom edge of the nano-ribbon.

Irradiating of circularly polarized light on a TI nano-ribbon which is doped by magnetic impurities, induces a phase transition from QAHI to NI phase. The points marked by $\alpha_2$ and $\alpha_4$ in Fig.~\ref{A1} are inside the NI phase region inducing by the right and left-handed circularly polarized light, respectively. Emerging normal insulator phase can be used for designing an {\it off}-state for electronic switch so that not only there are no bulk transport channels, but the edge modes also disappear. As an application for this region of the phase diagram, the nano-ribbon shown in Figs.~\ref{fig3} b1 and d1, results in a nullified current. The local current distribution drawn in Figs.~\ref{fig3} b2 and d2 demonstrates that for nano-ribbons longer than a penetration length, no current can pass through the nano-junction. 

The current-voltage characteristic curves for all the above cases of the mass-term step junctions are summarized in Fig.~\ref{fig3} e. The I-V curve shows an Ohmic behavior in two different types in which the slope of red line is 1, representing the contribution of only one edge in transport as depicted in Figs.~\ref{fig3} a, c. The {\it off} state is plotted with the black solid line as shown in Fig.~\ref{fig3} e.

For the sake of completeness, we investigated the simultaneous effect of SIA potential induced by perpendicular electric field and also illumination of circularly polarized light. Interestingly, as shown in Fig.~ \ref{A2}, the topological phase is QPHI phase for SIA potentials lower than a critical value in the dark mode. However, as shown in Ref.~\cite{dabiri2}, as a result of applying SIA potential, the Hamiltonian \ref{eq:firsthamil} does not break into pseudo-spin polarized parts and as a consequence, there is a mixture of psuedo-spin states. It is demonstrated that as far as the gap of pseudo-spin operator is opened~\cite{dabiri2}, the pseudo-spin chern number is still definable and the edge states are revived inside the band gap. However there is no pseudo-spin polarized edge state in TI nano-ribbon anymore. In fact, QPHI phase guaranties the existence of edge current at both edges of nano-ribbon, but in this time, there is a mixture of pseudo-spins flowing through each edge.
 
Now let us took three special points of the phase diagram represented in Fig.~\ref{A2}; $\beta_1$ as a QPHI phase, and $\beta_2, \beta_3$ as NI phases. In the dark mode, under application of SIA potential, the nano-junction sets in the parameters of the point $\beta_1$, guarantees the existence of non-polarized edge states, while the later points are used for the {\it off-} state of the switch. To realize the point $\beta_1$ in experiment, we suppose a perpendicular electric field applied only on the scattering region of the nano-ribbon in the absence of magnetic doping and also light illumination. As a result, as sketched in Fig.~\ref{fig4} a1, two edge currents containing a mixture of pseudo-spins are flowing through each edge. The local currents of the mixed states are marked by the black arrows. This proposed schematic view could be easily confirmed by the current distribution in Fig.~\ref{fig4} a2. Due to the lack of SIA potential applied in both electrodes, the input and output edge currents are pseudo-spin polarized. As a result, the I-V curve shows an Ohmic law in Fig.~\ref{fig4} d (blue line). 

 By turning on the circularly polarized light on the scattering region, a phase transition from QPHI to NI occurs and a gap is opened in the non-pseudo-spin polarized edge modes. As it is observed in the schematic view of Figs.~\ref{fig4} b1 and c1, pseudo-spin polarized currents are reflected into the source electrode and no current is passing through the nano-ribbon. It is equivalent to the off-state of the switch. Looking at the local current distributions in Figs. \ref{fig4} b2 and c2, demonstrates a quantum oscillation pattern for the current which tunnels through the bulk region of the nano-ribbon. The black line in I-V curve of Fig.~\ref{fig4} d, represents the off-state of the switch. As conclusion, we propose an electronic switching with off or on operation by turning on or off the circularly polarized light, respectively.

\section{Conclusion}\label{S5}
An electronic switch is designed by irradiating thin topological insulators realized on thin film of Bi$_2$ Se$_3$ and [(Bi,Sb)$_2$ Te$_3$] family at high frequency regime. The characteristic I-V curve of a nano-ribbon of TI thin film connected to two dark electrodes is studied by means of non-equilibrium Green's function formalism. In the scattering region, by illumination of the circularly polarized light, or a change in perpendicular magnetization and also an applied potential, one can control the mass term of the system and design a switcher. Thanks to the pseudo-spin polarized Hamiltonian in the absence of SIA potential, we presented the first proposition of the switcher based on the edge current through the TI's nano-ribbon. In this case, the scattering region is departed into two regions, each region is selectively shined by the LCP or RCP light. The polarization of the light gives the possibility of determining on which edge the current is demanding to pass. The off-state of the current is provided by illuminating each part by opposite circularly polarized light simultaneously.
 
 In the second set-up, by applying $M_z$, nano-ribbon is set in QAHI phase giving rise to the edge current. By turning on the LCP or RCP light, current through the edges is cut off. Finally, the last version of switcher, is operating by application of a perpendicular potential which leads to QPHI phase. The edge current in this case, is again cut off by illuminating the LCP or RCP light.

\appendix 
\section{Phase Diagrams}\label{A}
\begin{figure}
\includegraphics[width=0.9 \linewidth]{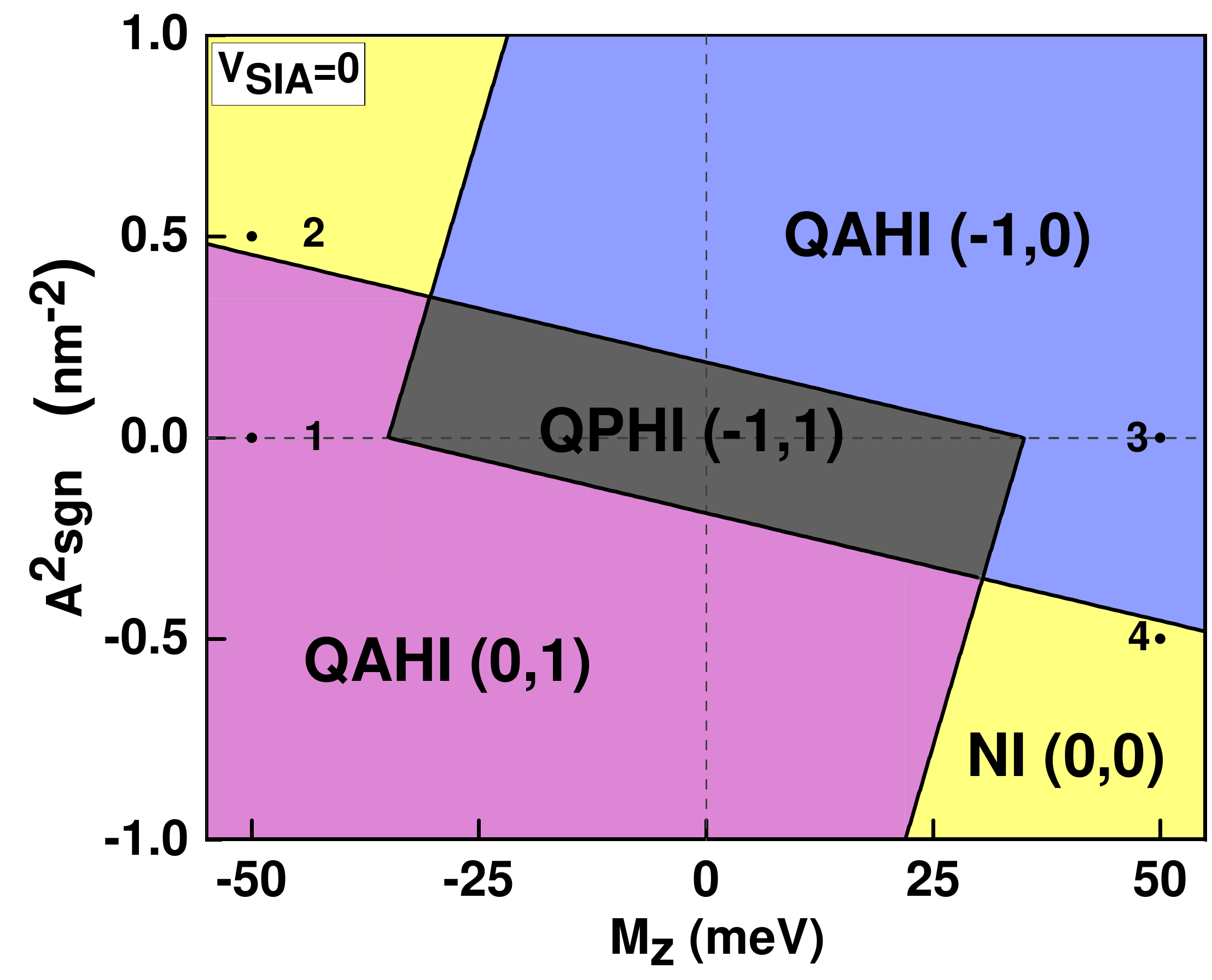} 
\caption{Phase diagram of a magnetically doped thin topological insulator which is irradiated by circularly polarized light.
Four distinct phases with specific pseudo-spin Chern numbers (C+, C-) as QPHI (-1,1), QAHI (-1,0), QAHI (0,1), NI (0,0) are achievable by different amount of magnetization and light intensity.} 
\label{A1}
\end{figure}

 \begin{figure}
\includegraphics[width=0.9 \linewidth]{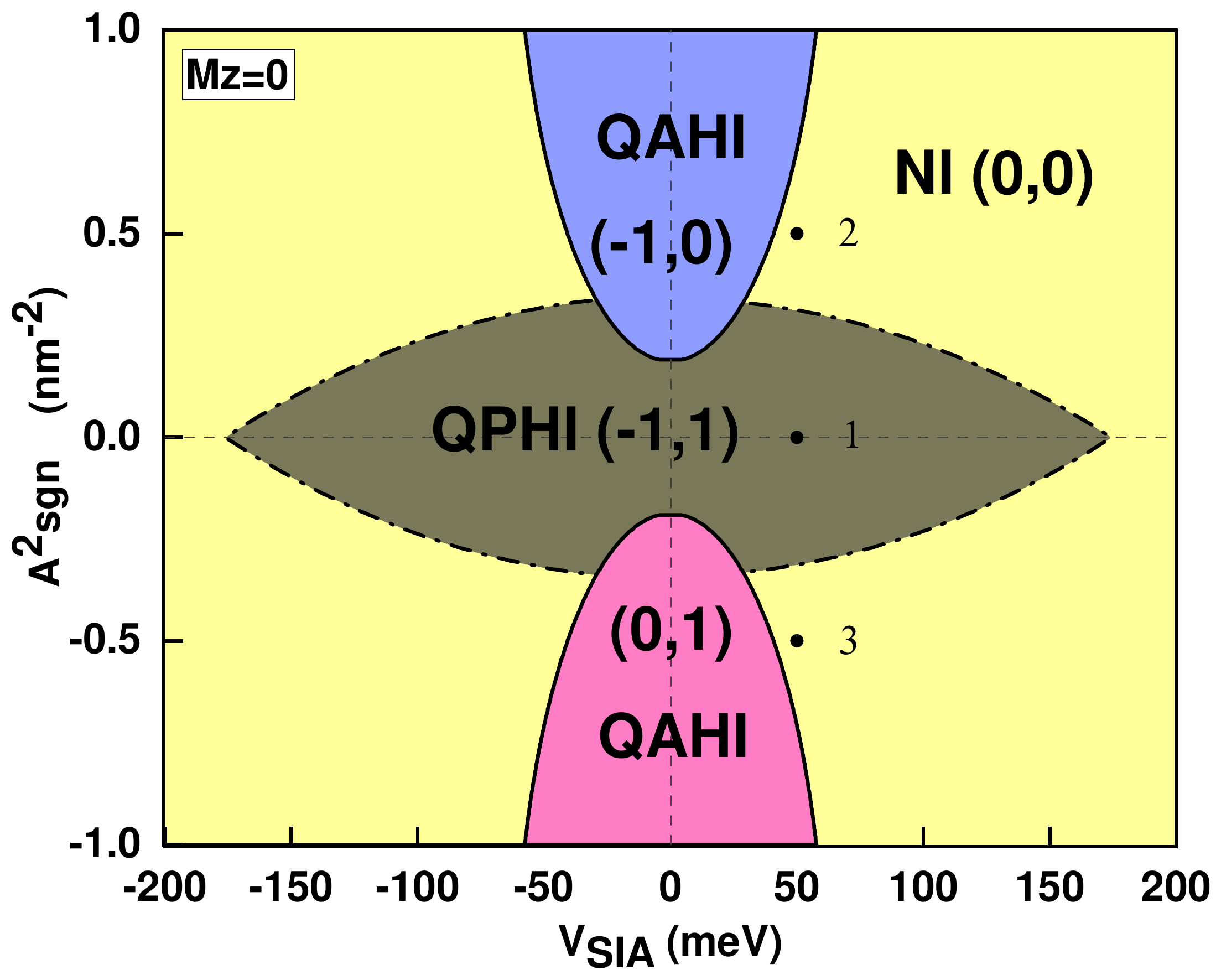} 
\caption{Phase diagram of a thin topological insulator irradiated by circularly polarized light and applied perpendicular electric field simultaneously. Four distinct phases with specific pseudo-spin Chern numbers (C+, C-) as QPHI (-1,1), QAHI (-1,0), QAHI (0,1), NI (0,0) are achievable by different amount of $V_{SIA}$ induced by electric field and light intensity. Solid (dashed) curves are related to the gap closing at (out of) the $\Gamma$ point.}
\label{A2}
\end{figure}

 
Two decoupled pseudospin Hamiltonians given by Eq.~\ref{Photo_Hamiltonian} in the absence of SIA, show several topological phases depending on the experimentally tuning parameters. In both absence and presence of magnetic impurities, one can tune $m_{\Omega}$ and modify the topological phases by changing the intensity, frequency, and polarization of light. 

Taking $M_{z}$ in account, the solid lines in phase diagram Fig.~\ref{A1} arising from $m_{\Omega}=-M_{z}+\Delta_0^{\prime}$ \cite{dabiri1}, divide the $M_{z}-A^{2}$ plane into several distinct topological phases denoted by gray region as QPHI (-1,1) with two $\alpha_{z}=-1$ and $\alpha_{z}=+1$ fully pseudo-spin polarized edge states , blue region as QAHI (-1,0) with one $\alpha_z=+1$ fully pseudo-spin polarized edge state, pink region as QAHI (0,1) with one $\alpha_{z}=-1$ fully pseudo-spin polarized edge state, yellow region as NI (0,0) fully gaped and without any edge state. $ \alpha$ points linked to the Fig. \ref{fig3} show the parameters for each configuration.

As mentioned before, the Hamiltonian \ref{Photo_Hamiltonian2} doesn’t commute with the pseudo-spin operator $\tau_z$ . Although the pseudo-spin Chern number is still definable, the quantum number $\alpha_z$ isn't invariant when $V_{SIA}\neq0$ .
Taking  $V_{SIA}$ induced by electric field in account, this time, the gap closing  occurs at the $\Gamma$ point ($k=0$) and $k^2=-\frac{\Delta'_0}{\Delta_1}-2A'm>0$ with the phase boundaries; the solid lines in phase diagram Fig. \ref{A2} when
\begin{equation}
V_{SIA}^2=m^2 -{\Delta'}_0^2
\end{equation}
in which $m=m_z+m_\Omega$ and the dash lines when\cite{dabiri2}. 
\begin{equation}
V_{SIA}^2=\frac{(-1+4A'^2\Delta_1^2)}{\Delta_1}(-m^2\Delta_1+\hbar^2 v_f^2(\Delta'_0+2A'm\Delta_1)) 
\end{equation}
Now there are still but differently located four distinct topological phases depicted by QPHI (-1,1), QAHI (-1,0), QAHI (0,1) and NI (0,0), in $V_{SIA}$-A2 plane but with mixed pseudo-spin states this time.

\bibliographystyle{apsrev4-2} 
\bibliography{aps} 

\end{document}